\documentclass{kluwer}
\usepackage{graphicx}
\usepackage{times}

\newcommand{\apj}{{\it Astrophys. J.}}
\newcommand{\apjl}{{\it Astrophys. J. Letters}}

\newcommand{\aap}{{\it Astron. \& Astrophys.}}
\newcommand{\aj}{{\it Astron. J.}}
\newcommand{\pasj}{{\it PASJ}}

\begin{document}
\begin{article}
\begin{opening}
\title{The  Apsidal Antialignment of  the  HD 82943  System
\footnote{Contribution to Proceedings for  IAU Colloquium No.189,
Astrophysical  Tides: Effects in  the  Solar and Exoplanetary
Systems, Nanjing, September 16-20, 2002}}
\author{Jianghui \surname{JI$^{1,3}$} \email{jijh@pmo.ac.cn} }
\author{Hiroshi  \surname{Kinoshita$^{4}$}}
\author{Lin      \surname{LIU}$^{2,3}$}
\author{Guangyu  \surname{LI}$^{1,3}$}
\author{Hiroshi  \surname{Nakai}$^{4}$}
\institute{
1.Purple  Mountain  Observatory, Nanjing 210008,China
\\2.Department of Astronomy, Nanjing University, Nanjing  210093,China
\\3.National Astronomical Observatory, Beijing 100012, China
\\4.National Astronomical Observatory, Mitaka, Tokyo
181-8588, Japan}
\runningauthor{Ji et al.}
\runningtitle{Apsidal Antialignment of the  HD 82943 System}
\begin{abstract}
We perform numerical simulations to explore the dynamical
evolution of the HD 82943 planetary system. By simulating diverse
planetary configurations, we find two mechanisms of stabilizing
the system: the 2:1 mean motion resonance between the two planets
can act as the first mechanism  for all  stable orbits. The second
mechanism is a dynamical antialignment of the apsidal lines of the
orbiting planets, which implies that the difference of the
periastron longitudes $\theta_{3}$ librates about $180^{\circ}$ in
the simulations. We also use a semi-analytical model to explain
the numerical results for the system under study.
\end{abstract}
\end{opening}

\section{Introduction}
At present, more than one hundred giant extrasolar planets have
been discovered in Doppler surveys of solar-type stars (Butler et
al., 2001, 2003), among which there are 10 multiple-planet
systems, including eight two-planet systems (HD 82943, GJ 876,  HD
168443, HD 74156, 47 Uma, HD 37124, HD 38529 and HD 12661) and two
three-planet systems (55 Cnc and $\upsilon$ And). The observations
indicate the Mean Motion Resonance (MMR) frequently occurs  for
the planets of the multiple-planet systems: the two planets of the
GJ 876 (Laughlin and Chambers, 2001; Kinoshita and Nakai, 2001;
Snellgrove, Papaloizou and Nelson, 2001; Lee and Peale, 2002; Ji,
Li and Liu, 2002) and HD 82943 (Gozdziewski and Maciejewski, 2001;
Beauge, Ferraz-Mello and Michtchenko, 2002; Ji and Kinoshita, in
preparation) are respectively in the 2:1 MMR, the inner two
companions of the 55 Cnc (Marcy et al., 2002; Fischer et al.,
2003; Ji et al., 2003) is in the 3:1 MMR and 47 Uma (Laughlin,
Chambers and Fischer, 2002) is close to a 7:3 commensurability,
which inspire us to study the resonant configurations of the
planetary systems.

We utilized N-body codes (Ji et al., 2002) to perform the
numerical integrations for the HD 82943 system with RKF7(8)
(Fehlberg, 1968). Moreover, we also used symplectic integrators
(Wisdom and Holman, 1991) to examine the same orbits to assure the
results for some cases.  In the simulations, the mass of the
central star is adopted to be 1.05 $M_{\odot}$, and those of the
two planets (HD 82943b, HD 82943c ) are respectively 1.63 and 0.88
$M_{Jup}$ under the assumption of $ \sin i = 1$.  Next, we
introduce the scheme to generate the initial six orbital elements
(semi-major axis $a$ , eccentricity $e$ , inclination $i$, nodal
longitude $\Omega $, apsidal argument $\omega $ and mean anomaly
$M$) for each planet. Throughout the paper, let us bear in mind
the hypothesis that the two planets are always considered to be
coplanar. For all the orbits, we assume that the semi-major axes
of the two planets always start at 1.16 AU and 0.73 AU (see Table
1)\footnote{The data are partly adopted from
http://cfa-www.harvard.edu/planets/encycl.html,
http://obswww.unige.ch/$\sim$udry/planet/hd82943syst.html}. We
take their initial eccentricities $e_{0}$ to be centered
respectively at 0.41 and 0.54, and randomly displaced by the
measuring error $\Delta e$.  And the initial arguments of
periastron are treated in similar way. The remaining two angles of
nodal longitudes and mean anomalies are randomly chosen between
$0^{\circ}$ and $360^{\circ}$. Thus, 100 pairs of coplanar orbits
are prepared for the integration. In this paper, we aim to study
the dynamical behavior of the HD 82943 system by exploring diverse
configurations in the neighborhood  of the best-fit solutions,
further attempt to discover the possible stabilizing mechanisms of
maintaining this system.

\begin{table}[ht]
\caption[]{The parameters of the HD 82943 planetary system}
\begin{tabular}{lll}
\hline Parameter
& HD 82943b &HD 82943c \\
\hline
$M$sin$i$($M_{Jup}$)       & 1.63  & 0.88  \\
Orbital period $P$(days)   & 444.6 & 221.6 \\
$a$(AU)                    & 1.16  & 0.73  \\
Eccentricity $e$           & 0.41  & 0.54  \\
$\Delta e$                 & 0.08  & 0.05  \\
$\omega$(deg)              & 117.8 & 138.0 \\
$\Delta\omega$(deg)        & 3.4   & 10.2  \\
\hline
\end{tabular}
\end{table}

\section{Apsidal antialignment}
In the numerical investigations, our goal is to understand the
characteristic of the secular behavior of the HD 82943 planetary
system. By using the aforementioned initial orbits, each
integration was executed for 10 Myr. As a result, we find that
$75\%$ of the orbits are unstable for the timescale of $10^{5}$ yr
and only $5\%$ of the experiments survived for $10^{7}$ yr. It is
noteworthy that all the stable cases are involved in the 2:1 MMR
and that the stability of a system is obviously sensitive to its
initial planetary configuration when we fix the masses of the
planets. In this paper, we discuss a stable configuration in
association with the 2:1 MMR and apsidal resonance.

At first, we introduce the lowest order critical arguments
$\theta_{1}$ and $\theta_{2}$ for the 2:1 MMR of the HD 82943
system
\begin{equation}
\label{eq1} \quad\quad\theta_{1} = \lambda _{1} - 2\lambda _{2} +
\tilde {\omega} _{1},
\end{equation}
\begin{equation}
\label{eq2} \quad\quad\theta_{2} = \lambda _{1} - 2\lambda _{2} +
\tilde {\omega} _{2},
\end{equation}
\noindent where $\lambda _{1} $, $\lambda _{2} $ are ,
respectively, the mean longitudes of the inner and outer planets,
and $\tilde {\omega} _{1} $,  $\tilde {\omega} _{2}$ denote their
apsidal longitudes  respectively (subscript 1 for the inner
planet; 2, the outer planet).  Figure 1a shows that one of the
resonant arguments $\theta_{2}$ librates about $0^{\circ}$ for the
timescale of 10 Myr (in fact, $\theta_{1}$ also librates around
$180^{\circ}$ for the same timescale), indicating the two planets
of the system are now locked into the 2:1 MMR. On the basis of the
best-fit  solutions  by  Laughlin and Chambers (2001) for  GJ
876, Lee and Peale (2002) found that $\theta_{1}$ and $\theta_{2}$
both librate  about $0^{\circ}$ with  small  amplitudes.
Moreover, they pointed out that the differential planet migration
due to planet-nebular interaction could give rise to capture into
the 2:1 MMR. From the Laughlin-Chambers solutions, one can see
that the outer planet of the GJ 876 moves on a near circular
orbit. By contrast, the HD 82943 system differs in that both of
the planets occupy high eccentricities at present-day. Hence, the
origin of the 2:1 resonance for the HD 82943 surrounded by two
massive eccentric planets should be explained by a new mechanism
(S. Peale, private communication), which may be quite different
from the resonant origin of the GJ 876.

Most important of all, we find that the periodic coplanar orbits
of the HD 82943 apparently cross during the secular orbital
evolution because of high eccentricities of two planets. To the
best of our knowledge, such kinds of the stable orbits are never
reported before. One should make clear that what kinds of the
mechanisms make  the system stable for tens of millions of years.
However, we notice that the planets of the studied system
simultaneously undergo apsidal antialignment . The relative
apsidal longitude $\theta_{3}$ is denoted by
\begin{equation}
\label{eq7} \theta_{3} = \theta_{1} - \theta_{2}=\tilde\omega_{1}-
\tilde\omega_{2} .
\end{equation}
\noindent From  Figure 1b, we see that $\theta_{3}$ librates about
$180^{\circ}$ with an amplitude of $\pm 30^{\circ}$, conversely,
Lee and Peale (2002) found that $\theta_{3}$ librates about
$0^{\circ}$ for the GJ 876, which indicates the alignment of the
apsidal lines. However, it is the first time to observe the
antialignment configuration for the HD 82943 system.  Recently,
the antialignment for the HD 12661 was independently confirmed  by
Lee and Peale (2003a) and Gozdziewski (2003). The librations of
the relative apsidal longitude were also discovered in $\upsilon$
And (Kinoshita and Nakai, 2000; Chiang, Tabachnik and Tremaine,
2001) and 47 Uma (Laughlin et al., 2002). Besides these symmetric
apsidal resonance , Ji  et al. (2003) found the asymmetric apsidal
librations about $250^{\circ}$ or $110^{\circ}$  for the 55 Cnc,
and the asymmetric configurations were also suggested by Lee and
Peale (2003b) and  Beauge et al. (2002). As for $\upsilon$ And,
Kinoshita and Nakai (2000) reported the mechanism of the apsidal
alignment by using the linear secular perturbation theory. Now it
is believed that the apsidal resonance significantly plays a part
in stabilizing the  multiple-planet systems. In the following, we
utilize a semi-analytical model to study the dynamics near the
apsidal resonance and then compare the analytical results with the
numerical outcomes.

The Hamiltonian for the coplanar case (Kinoshita and Nakai, 2002)
is
\begin{equation}
\label{eq8} F=F(a_1, a_2, e_1, e_2, \tilde {\omega} _{1}, \tilde
{\omega} _{2}, \lambda_1, \lambda_2).
\end{equation}
In order to keep the Hamiltonian form, we should use Jacobi
coordinates or canonical heliocentric coordinates to study the
system. As the indirect part of the Hamiltonian does not
contribute to the secular part, we simply take the direct part. As
aforementioned, the orbits of two planets intersect each other,
the usual analytical development method of the main part can not
be applicable to this case,  then we adopt the original form of
the main part of the disturbing function and numerically evaluate
it.  Furthermore, by eliminating short-periodic terms, the new
Hamiltonian reads:
\begin{equation}
\label{eq9} F^*=F^*(a_1, a_2, e_1, e_2, \theta_2, \theta_3).
\end{equation}
The degrees of freedom of the new Hamiltonian (\ref{eq9}) are
reduced from four to two. However this Hamiltonian is still not
integrable. As the amplitude of the critical argument $\theta_2$
is not so large, we assume $\theta_2$=0. From this assumption, the
semi-major axes $a_1$ and $a_2$ are derived as constant and we use
$a_1$=0.73 AU and $a_2$=1.16 AU in the following discussions.
Correspondingly, the degrees of freedom of the new Hamiltonian are
reduced from two to one and the new Hamiltonian takes the
following form:
\begin{equation}
\label{eq10} F^*=F^*(e_1, e_2, \theta_3).
\end{equation}
\noindent We eliminate $e_2$ in equation (\ref{eq10}) with the
conservation of the angular momentum $H$ , then we have
\begin{equation}
\label{eq12} F^*=F^*(e_1, \theta_3, H) .
\end{equation}
\noindent Thus we can plot the level curves of the Hamiltonian and
understand the global behavior of $e_1$ and $\theta_3$. We draw
the contour map of the Hamiltonian (\ref{eq12}) by taking
$\theta_3$ as the horizontal axis and $e_1$ or $e_2$ as the
vertical axis with the parameter $H$,  which is determined from
the initial conditions. Figure 2 shows the contour map of the
Hamiltonian (\ref{eq12}) given by thin lines. Because the
numerical solution includes the short-periodic terms, the
numerically averaged solution represented by broad thick lines is
shown in Figure 2. The figure shows a good agreement between the
numerical solution and the semi-analytical secular solution. The
eccentricities of both planets are well restricted because of the
libration of $\theta_3$ around $180^{\circ}$. Therefore, we may
safely conclude that the stability of the HD 82943 system that is
related to the above coplanar crossing orbits can be
simultaneously sustained by two mechanisms-the 2:1 MMR and the
apsidal antialignment.

\section{Conclusions}
In final, we summarize some conclusions: we carried out the
numerical simulations of the HD 82943 system for $10^{7}$ yr. In
the simulations,  we find that all of the stable orbits are
associated with the 2:1 MMR. In particular, we underline that a
stable case of the coplanar crossing orbits is not only in a 2:1
MMR, but also experiences the apsidal resonance during the secular
orbital evolution. The presentence of the 2:1 MMR reveals the fact
that the resonance can act as one of the mechanisms of stabilizing
the HD 82943 system that harbors two eccentric Jupiter-like
companions. Additionally, the second mechanism is a dynamical
antialignment of the axes of the orbits of the two planets,
showing that $\theta_{3}$ librates about $180^{\circ}$ for the HD
82943 system.  The $\theta_{3}$-libration  makes it possible that
the massive planets in the mean motion resonance avoid frequent
close encounters. Furthermore, we point out that to date the
discovered multiple-planet systems in the MMR (HD 82943, GJ 876,
HD 12661, 47 Uma , 55 Cnc ) are always followed by apsidal
resonance and this can be easily understood that the apsidal
libration can result from the librations of the mean motion
resonant arguments. This apsidal libration is also responsible for
the stability of nonresonant systems, such as $\upsilon$ And and
other cases.

\centerline{\bf Acknowledgments} We thank Stan Peale for the
informative discussions. We also thank  Beauge C. and Ferraz-Mello
S. for their insightful comments that helped improve the
manuscript. We are grateful to the referee for language
corrections.This work is financially supported by the National
Natural Science Foundations of China (Grants 10203005, 10173006,
10233020), and the Foundation of Minor Planets of Purple Mountain
Observatory.

\clearpage

\begin{figure}
\caption{(a) The
upper panel shows that one of the resonant arguments of
$\theta_{2}$ librates about $0^{\circ}$ for 10 Myr, indicating the
two planets of the HD 82943 system are now locked into the 2:1
MMR. (b) The lower panel  displays that the argument of
$\theta_{3}$ librates about $180^{\circ}$ for the timescale of 10
Myr,  with an amplitude of $\pm 30^{\circ}$ for the system, which
implies the antialignment for this system. Note: the initial
conditions for numerical integration: $a_{1}=0.73$,
$e_{1}=0.5440$, $\Omega_{1}=8.01^{\circ}$,
$\omega_{1}=132.55^{\circ}$, $M_{1}=267.66^{\circ}$. $a_{2}=1.16$,
$e_{2}=0.4628$, $\Omega_{2}=223.94^{\circ}$,
$\omega_{2}=117.89^{\circ}$, $M_{2}=214.50^{\circ}$. }
\label{Fig1}
\end{figure}

\clearpage
\begin{figure}  
\caption{The
equi-Hamiltonian curves for the eccentricities of two planets
versus $\theta_{3}$. The thin lines are computed from numerically
averaged Hamiltonian by assuming that the planetary system is in
exact 2:1 MMR. The broad thick line in the libration region around
$\theta_{3}=180^{\circ}$ shows the solution obtained by numerical
integration, which suggests a good agreement with the
semi-analytical method for the case of the libration. }
\label{Fig2}
\end{figure}

\end{article}
\end{document}